\documentclass[twocolumn,showpacs,preprintnumbers,amsmath,amssymb,amsfonts,superscriptaddress,floatfix,prl,aps]{revtex4-2}
\usepackage{CJK}

\usepackage[paperwidth=210mm,paperheight=297mm,centering,hmargin=2cm,tmargin=1.6cm,bmargin=3.cm]{geometry}
\usepackage{booktabs}
\usepackage{subfigure}
\usepackage{graphicx}
\usepackage{amsfonts}
\usepackage{amssymb}
\usepackage{amsmath}
\usepackage{overpic}
\usepackage{siunitx}
\usepackage{braket}
\usepackage{xcolor}
\usepackage{cancel}
\usepackage{fancyhdr}
\usepackage[utf8]{inputenc}
\usepackage{natbib}
\usepackage{lipsum}
\usepackage[T1]{fontenc}
\usepackage[newcommands]{ragged2e}
\usepackage[font=small,labelfont=bf,justification=Justified]{caption}
\usepackage{xcolor}

\newcommand{\be}{\begin{equation}}
\newcommand{\ee}{\end{equation}}
\newcommand{\bea}{\begin{eqnarray}}
\newcommand{\eea}{\end{eqnarray}}
\makeatletter
\makeatletter \renewcommand{\fnum@figure}{{\bf{\figurename~\thefigure}}}

\makeatother

\begin{document}

\title{ {BKT phase transition in nanoporous films of superconducting NbN}}

\author{A. Verma}
\thanks{These authors contributed equally}
\affiliation{School of Physical Sciences, UM-DAE Centre for Excellence in Basic Sciences, University of Mumbai, Kalina Campus, Santacruz (E), Mumbai 400098, India}
\author{R. Vedin}
\thanks{These authors contributed equally}
\affiliation{Department of   Physics, The Royal Institute of Technology, Stockholm SE-10691, Sweden}
\author{J. Jesudasan}
\affiliation{Department of Condensed Matter Physics and Materials Science, Tata Institute of Fundamental Research, Dr. Homi Bhabha Road, Colaba, Mumbai 400005, India}
\author{J. Lidmar}
\email{jlidmar@kth.se}
\affiliation{Department of   Physics, The Royal Institute of Technology, Stockholm SE-10691, Sweden}
\author{I. Maccari} 
\email{imaccari@phys.ethz.ch}
\affiliation{Department of   Physics, The Royal Institute of Technology, Stockholm SE-10691, Sweden}
\affiliation{Institute for Theoretical Physics, ETH Zurich, CH-8093 Zurich, Switzerland}
\author{S. Bose}
\email{sangita@cbs.ac.in}
\affiliation{School of Physical Sciences, UM-DAE Centre for Excellence in Basic Sciences, University of Mumbai, Kalina Campus, Santacruz (E), Mumbai 400098, India}
\date{\today}

\begin{abstract}
We present a study of the Berezinskii-Kosterlitz-Thouless (BKT) transition in mildly disordered NbN nanoporous (NP) films. The measured superfluid stiffness, $J_s$, is found to be much lower than that predicted by considering the reduction in the geometric area. This effect is also reproduced theoretically via Monte Carlo simulations on a 2D XY model with different nanopore geometries. For a 5 nm thick NP film, a distinct BKT transition is observed. BKT renormalization group flow equations, incorporating the broadening in $J_s$ due to the presence of inhomogeneities, are used to fit the experimental data. From this analysis we see that both $J_s$ and the vortex core energy, $\mu$, decrease in the presence of nanopores.
Our results show that nanopore geometries effectively enhance the 2D nature of the films, thereby increasing the parameter space to explore BKT physics in superconducting films.
\end{abstract} 

\maketitle

In two-dimensional (2D) superconductors while the emergence of a finite superconducting (SC) gap
occurs at the Bardeen-Cooper-Schrieffer (BCS) critical temperature, $T_c$, a global phase coherence among Cooper pairs may be established at a lower temperature~\cite{Goldman2012,Raychaudhuri2022},  $T_{BKT}<T_c$, associated with the Berezinskii-Kosterlitz-Thouless (BKT) transition~\cite {berezinskyDestructionLongrangeOrder1972,Kosterlitz1973, Kosterlitz1974, Nelson1977, Beasley1979, Epstein1981}. 
At the BKT critical temperature, the phase coherence is suddenly lost as a result of the proliferation of topological excitations of the phase, vortices, tightly bound in pairs at low temperatures. The unbinding of vortex-antivortex pairs has its most distinct experimental manifestation in the discontinuous jump to zero of the superfluid stiffness, $J_s$, at $T=T^+_{BKT}$.
In several 2D systems, ranging from conventional superconductors such as Nb, NbN, TiN, NbTiN
 to high-$T_c$ cuprate superconductors and novel superconductors like the boron-doped diamond, Weyl semi-metals as PtBi$_{2}$, and 2D heterostructures as LaAlO$_3$/SrTiO$_3$,AlO$_x$/KTaO$_3$(111),  
 the BKT transition has been probed experimentally~\cite{Lemberger2007,Kamlapure2010,Baturina2012,Yong2013,Burdastyh2020,Baity2016,Hetel2007, Mallik2022,Ojha2023,Jarjour2023,Veyrat2023,Coleman2024}. 
 These studies have employed various techniques, including dc transport, penetration depth or kinetic inductance measurements~\cite{Beasley1979,Lemberger2007,Mondol2011,Baturina2012,Mallik2022,Weitzel2023}. 
In 2D superconductors, the BKT transition only occurs if the Pearl length $\Lambda = \lambda^2/t$~\cite{Pearl1964} is large compared with the system size, where $\lambda$ is the magnetic penetration depth and $t$ is the thickness of the film.
Even when this condition is met in very thin films, clear experimental signatures can only be observed if the superfluid stiffness, $J_s$ is smaller than the SC gap, $|\Delta|$, ensuring that $T_{BKT}$ is well separated from $T_c$~\cite{Weitzel2023}.
 At the same time, the temperature dependence of 
 $J_s(T)$ and its expected jump at $T_{BKT}$ crucially depends on the value of the vortex-core energy, $\mu$~\cite{Benfatto2007, Mondol2011}. That is
the energy cost to nucleate a vortex, being~\cite{Bruun2001} $\mu \approx \pi \xi^2 \epsilon_{cond}$, where the coherence length, $\xi$, is taken to be the linear size of the vortex core and $\epsilon_{cond} \propto |\Delta|^2$ is the condensation energy. Small values of $\mu$ 
 renormalize $J_s$ 
 to lower values, reducing the value of $T_{BKT}$ further and, hence, its distance from $T_c$.
Low values of $J_s$ and $\mu$, are generally found in disordered or extremely thin SC films making them the best candidates to explore BKT physics. 
 However, in these systems, the high disorder level —arising from atomic-level point defects, intrinsic or emergent granularity ~\cite{Ghosal1998, Mondol2011,Sacape2008,Carbillet2016,Chen2017}—
 may induce strong inhomogeneity in the SC order parameter which in turn can lead to a significant broadening of the superfluid density jump~\cite{Benfatto2009,Maccari2017,Maccari2018,Maccari2019,Venditti2019}. The smearing of the expected sharp jump of $J_s$ at $T_{BKT}$ greatly hinders a clear identification of the BKT transition. 
 The question which naturally follows is -  Is it possible to reduce $J_s$ and $\mu$ simultaneously without making the superconductors extremely thin or highly disordered? A restricted geometry arising from nano-patterning a superconductor may be able to reduce $J_s$ and $\mu$ simultaneously without broadening the BKT transition.
\par
There have been several studies in the past where nano-patterning of the SC thin film has been attempted. They have been extensively used to investigate the phenomena of vortex matching effects (VME) and the role of the nanopores on vortex dynamics ~\cite{Moshchalkov2000,Welp2002,Kumar2013, Kumar2015, Bose2023}. Few reports also exist where these nano-patterned films have been used to study the effect of disorder on the superconductor-insulator transitions (SIT) which get enhanced due to the restricted geometry ~\cite{Baturina2011,Hollen2013}. There have also been a few studies to explore the BKT physics in these nano-patterned films forming a 2D array of Josephson junctions. It was observed that when the Josephson coupling energy, $E_J$, becomes larger than 
the charging energy, $E_C$, the BKT transition becomes possible~\cite{Baturina2011, Goldman2012}.
\par
In this paper, we present the study of BKT transition in moderately disordered SC NbN films in the presence of periodic arrays of nanopores.  Nanoporous (NP) films, with a thickness (t) between \ {15} {nm} and \ {5} {nm}, are studied by transport and penetration depth measurements. 
Our results show that the nanopores reduce $J_s$ by an amount which is larger than the expected decrease due to the restricted geometry (in films of comparable disorder and of any thickness).
As a consequence, 
for the thinnest NP films ($\approx {5} {nm}$), the BKT signatures become more distinct and the two temperatures $T_{BKT}$ and $T_c$ can be clearly identified.  
Moreover, signatures of the jump in $J_s$ at $T_{BKT}$ are also observed in NP films with thicknesses as large as\ {15} {nm}. 

To shed light on the experimental results, we study a 2D XY model in the presence of regular arrays of holes via Monte Carlo simulations. Our numerical results demonstrate how the presence of nanopores can reduce both $J_s$ and $\mu$ compared to the continuous case. 
Finally, BKT renormalization group flow equations, which incorporate the broadening in $J_s$ due to the presence of inhomogeneities, 
nicely fit the experimental data, showing a decrease in $\mu$  and an increase in $\mu$/$J_s$ in NP films. At the same time, this analysis reveals that the level of inhomogeneities, not renormalized by $J_s$, does not increase in NP films, very differently from what typically occurs by reducing the film thickness.
Our work shows that NP films are a good model system to explore BKT physics enabling the artificial tuning of the two energy scales at play, $J_s$ and $\mu$, without the drawback of increasing the disorder strength.

\begin{figure}[h!]
    \centering
    \includegraphics[width=1.05\linewidth, scale=1.1]{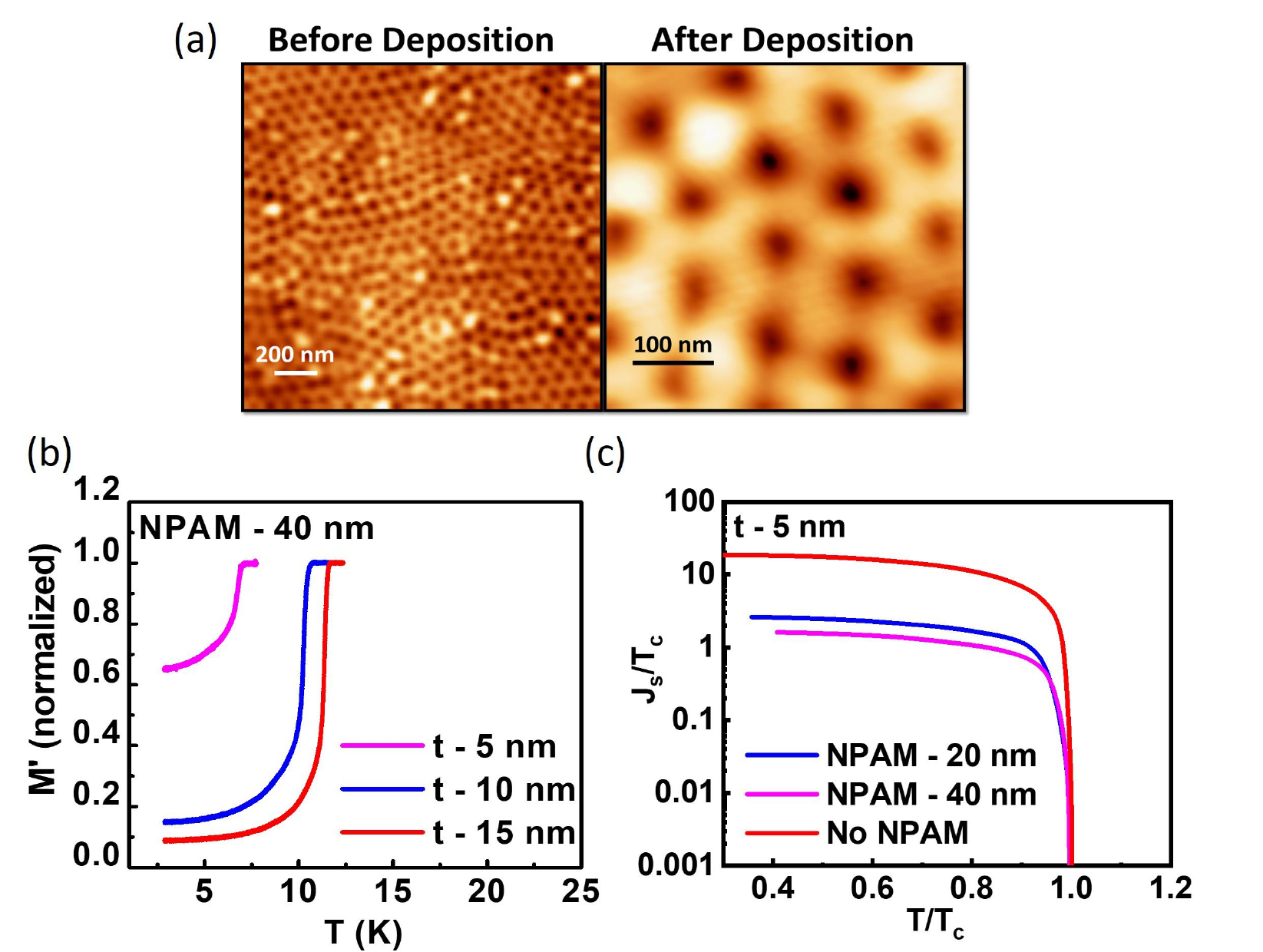}
    \caption{(a) AFM image of the NPAM with pore diameter of 40 nm before (left panel) and after (right panel) deposition of NbN. (b) Real part of mutual inductance ($M{'}$) $\it {vs}$ T for films with thickness (t) varying between\ {15} and\ {5} {nm} on the NPAM with pore diameter of\ {40} {nm}. (c) Temperature variation of $J_s$ extracted from penetration depth measurements for 5 nm thick continuous and NP films with pore diameters of \ {40} {nm} and \ {20} {nm}. Both  $J_s$ and T are normalized with  $T_c$.}
    \label{fig:char}
\end{figure}

NbN thin films of varying thickness (t) were grown by reactive DC magnetron sputtering on free-standing nanoporous anodic aluminium oxide membranes (NPAM) obtained from InRedox. NPAMs with nominal hole diameters (d) of\ {40}{ nm} (Atomic force microscope, AFM image shown in left panel of Fig. \Ref{fig:char} (a)) and\ {20} {nm}  with pore periods of\ {100} {nm} and\ {50} {nm} respectively were used. For details related to film growth, \textcolor{blue}{see supplementary information} and references \cite{Chocka2008,Mondol2011}. AFM image of the NP NbN film grown on the NPAM with\ {40}{ nm} pore diameter is shown in the right panel of Fig.\ref {fig:char}(a) where NbN deposits along the walls of the NPAM.  For comparison, continuous films of similar thickness were also grown on Si(100) substrates. The $T_c$ of the films was determined through measurements of electrical resistivity (film grown in strip-line geometry) and diamagnetic shielding response (film grown in circular geometry), conducted in a Cryogenics, UK cryogen-free system, capable of reaching temperatures of \ {300} {mK} and magnetic fields up to \ {9} {T}. The diamagnetic shielding response and the penetration depth ($\lambda $) measurements were done using a custom-designed two-coil mutual inductance probe integrated with the cryogen-free system. The details of the set-up and the measurements are the same as in Ref ~\cite{Kumar2013, Gupta2020}. The real (imaginary) mutual inductance $M{'}$($''$) was measured as a function of temperature,
from which we extracted both $ \lambda $ and $J_s$ 
as explained in Ref~\cite{Gupta2020,Raychaudhuri2022}. The temperature dependence of $M{'}$ for films of $t$ ranging between\ {15} -\ {5} {nm} grown on NPAMs with diameter of\ {40} {nm} is shown in Fig. \Ref{fig:char}(b). Similar measurements on films of 
different thickness $t$, grown on NPAMs with pore-diameter of\ {20} {nm}, are shown in \textcolor{blue}{Fig.~S4 of the supplementary information}. As expected, the $T_c$ decreases with film thickness and does not depend substantially on the pore size. However, the $T_c$'s of the films on the NPAMs were lower when compared to continuous films. The $T_c$ of the\ {5} {nm} thick NP films was reasonably high and ranged between 6 - 7 K (for different growth cycles) which indicated that the films were mildly disordered (See \textcolor{blue}{see supplementary information, 
and Fig.~S1}  and ~\cite{Mondol2011,Kamlapure2010}).

\begin{figure}[h!]
    \centering
     \includegraphics[width=1.1\linewidth]{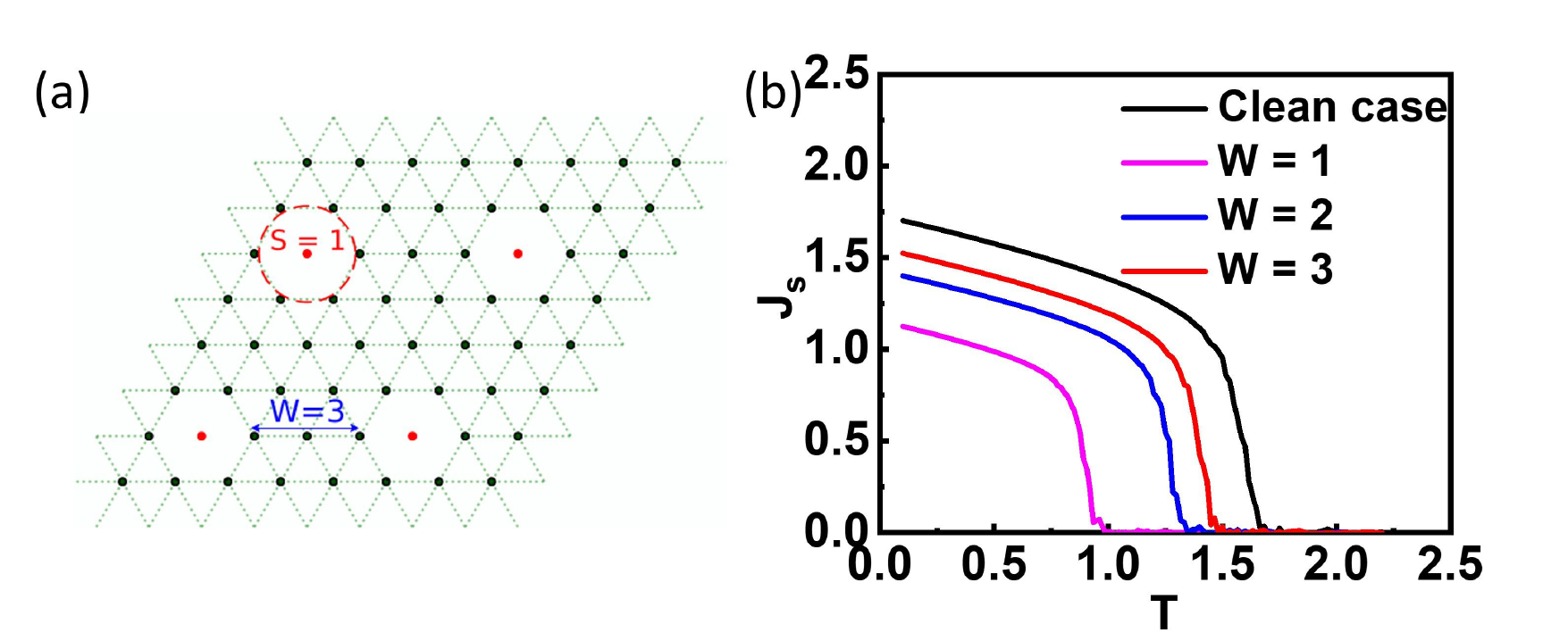}
    \caption{(a) Schematic illustration of the simulated lattice, red dots correspond to sites inside the pores while the green dashed lines represent the coupling $J$ between sites. (b) Temperature dependence of $J_s$ from Monte Carlo simulations of 2D XY model for fixed nanopore size (S = 1) and different inter-pore separation (W). $J_s$ and T are given in dimensionless units, and measured in terms of the XY model coupling $J$.}
    \label{fig:js}
\end{figure}


\begin{figure*}
 \centering
    \includegraphics[width = \textwidth]{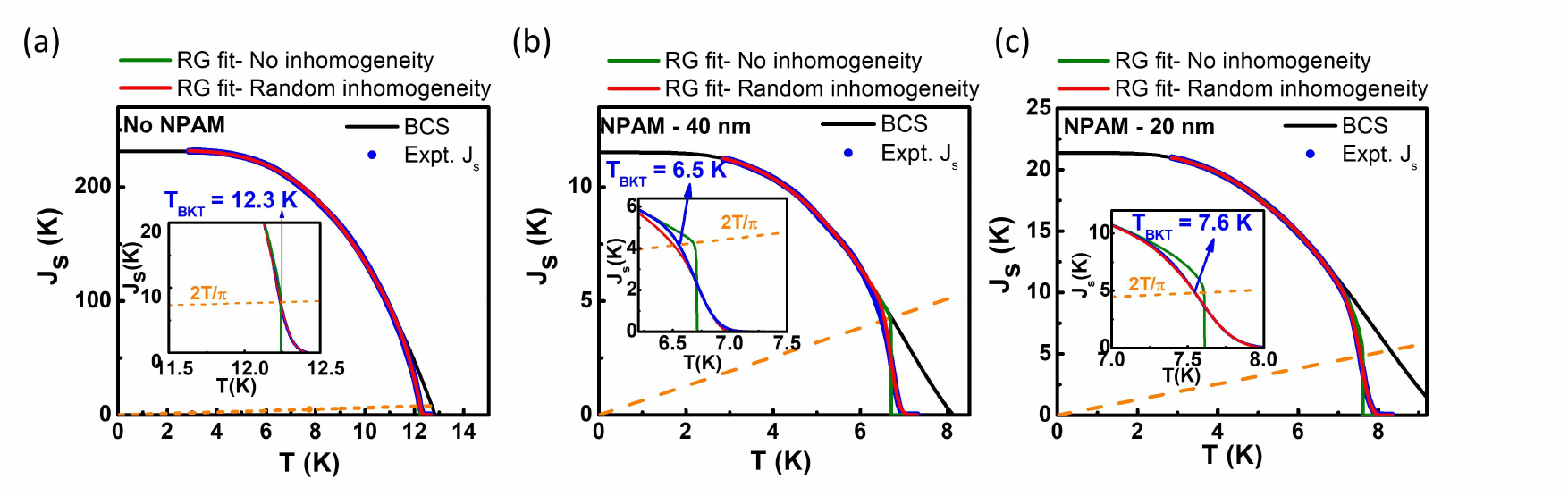}
    \caption{ (a) - (c) Temperature variation of  $J_s$ extracted from penetration depth measurements (blue solid symbols) for \ {5} {nm} thick continuous (without NPAM) and NP films with pore diameters of \ {40} {nm} and \ {20} {nm}. The black solid line is the BCS variation. The orange dashed line is the 2T/$\pi$ line. The green line shows the fit to the experimental data using the RG flow without including any inhomogeneity and the red solid line is the RG fit including random inhomogeneity. The insets show the zoom-in of the variation close to $T_c$ to identify $T_{BKT}$.}
    \label{fig:bkt}
\end{figure*}
\par
To see the effect of the nanopores on the superfluid stiffness, $J_s$ (normalized with $T_c$) is plotted as a function of temperature (also normalized with $T_c$) for the  \ {5} {nm} film (with and without NPAM) as shown in Fig. \Ref {fig:char} (c). A clear suppression is seen in the NP films. This suppression is higher than that expected due to the decrease in superfluid density caused by the change in the system geometry. To understand this better,  we performed Monte Carlo (MC) simulations on a 2D XY model and computed $J_s$ for different nanopore geometries, see schematic in Fig. \ref{fig:js}(a).
Each nanopore geometry is identified by $S$, the size of the nanopores corresponding to the number of lattice sites within each one, and $W$, the minimum distance between two neighbouring pores.  
We considered a triangular lattice geometry of $N = L \times L$ sites with periodic boundary conditions. The 2D XY Hamiltonian reads
\begin{equation}
    H_{XY}= -\sum_{\langle i, j\rangle} J_{i, j} \cos(\theta_i - \theta_j),
\end{equation}
where the sum is restricted to nearest neighbours sites $\langle i, j\rangle$, $\theta_i$ is the superconducting phase associated with the coarse-grained site $i$, and the presence of nanopores is implemented through spatial-dependent couplings $J_{i, j} = J n_i n_j$, such that $n_i = 0$ if the site $i$ belongs to the nanopore, and $n_i = 1$ if it belongs to the SC film. Different from previously studied XY models with quenched disordered couplings~\cite{leonel2003, wysin2005, erez2013, Maccari2016,  Maccari2017, Maccari2018, Maccari2019}, here the $\it{on-site}$ dilution is \emph{ordered} forming a periodic structure of non-SC islands. For each geometry,  system size, and temperature, we performed $5.5\times 10^4$ MC steps, consisting of 4 Metropolis local updates of the whole lattice followed by 4 microcanonical over-relaxation updates to speed up the thermalization. To reach equilibrium at lower temperatures, a simulated annealing procedure was implemented where the temperature is gradually reduced starting at a high temperature with an initial disordered state. The superfluid stiffness is computed as the linear response of the system to a uniform twist of the SC phase,  $\Theta_\nu$, along a given direction $\nu$, $J^\nu_s = \frac{2}{\sqrt{3} N_{SC}}  \frac{\partial^2F(\Theta_\nu)}{\partial \Theta_\nu^2}\Big|_{\Theta_\nu = 0}$, where $N_{SC}$ is the number of SC sites, $\frac{2}{\sqrt{3}}$ is a geometrical factor, and $F$ is the free energy of the system (\textcolor{blue}{see supplementary information} for more details). The resulting temperature dependence of $J^x_s \equiv J_s$ is shown in Fig.~\ref{fig:js} (b) for a linear lattice size, $L=108$, and nanopore geometries of pore size, $S=1$ and separations of $W=1,2,3$.  
The temperature dependence of $J_s$ for a nanopore geometry with larger pore sizes, $S=7$, is reported in Fig. S6 of the supplementary information. 
Our numerical simulations reveal that by decreasing the inter-pore separation, $J_s$ decreases and also gets suppressed when compared to that of the clean case, very similar to the experimental results (See Fig. ~\ref{fig:char} (c)). Note that, by definition, the computed superfluid stiffness is already renormalized by the effective superconducting density, i.e. number of SC sites, so its decrease cannot be simply ascribed to the change in the system geometry.

\par
Next, we identified and characterized
the BKT transition in the NP films. First, the I-V measurements carried out at temperatures $T$ below $T_c$ were analyzed based on the Nelson-Halperin scaling theory~\cite{Nelson1977}
to get the temperature variation of $J_s$ and consequently $T_{BKT}$. For details \textcolor{blue}{see supplementary information and Fig. S2}. For the\ {5} {nm} thick NP film, $T_{BKT}$ could be clearly identified over the smearing of the BKT jump associated with film inhomegeneity~\cite{Venditti2019, Benfatto2009}. However, what seems surprising is that for the film with thickness, t =\ {15} {nm}, larger than the coherence length, $\xi$ $\approx$ \ {5 - 7} {nm} \footnote{Note: $\xi$ $\approx$ \ {5 - 7} {nm}, as determined from, $\xi^2$ = $\frac {\phi_0}{2\pi H_{c2}}$, where, $\phi_0$ is the flux quantum and $H_{c2}$ is the upper critical field}, small yet visible signatures of BKT transition were observed. 
That may be explained by noting that the pore sizes of the NP films (40 and 20 nm) is larger than the thickness of the films and it can be 
considered as the effective vortex size. Furthermore, the observation of VME in these films indicates that these pores act as vortex-trapping sites.
Thus, it can be conjectured that the pore diameter is the relevant length scale dictating the vortex fluctuations in NP films. 
Alongside the I-V characteristics, the temperature dependence of the superfluid stiffness (in K) was also obtained from the penetration depth ($\lambda$) measurements via $J_s= \frac{\hbar^2 t}{4 k_B \mu_0 e^2 \lambda^2}$ \cite {Raychaudhuri2022}, where $\hbar$ is the Planck’s constant, $e$ the elementary charge, $k_B$ the Boltzmann’s constant and $\mu_0$ is the vacuum permeability.
The temperature variation of $J_s$ 
was first analysed via a standard BCS fit~\cite {Tinkham1996}, being
\begin{equation}
J_{BCS}(T) = J_s(0) \frac{\Delta(T)}{\Delta(0)} \tanh(\frac{\Delta(T)}{2 k_B T})
 \label{BCSfit}
\end{equation}
and 
\begin{equation}
\Delta(T)= \Delta(0) \left[ 1 - \frac{T^4}{3T_c^4}\right] \sqrt{1 - (T/T_c)^4}.
\end{equation}
In Fig.\Ref {fig:bkt}(a)-(c), the experimental data (blue solid circles) are shown along with the BCS fits (black solid line) for the\ {5} {nm} thick film, without and with NPAMs. Similar analysis was also done for films with thicknesses of\ {10} {nm} and\ {15} {nm} (\textcolor{blue}{see supplementary information, Fig. S3 and Fig. S4}). In the fits, $J_s$(0), $\Delta$(0) and $T_c$ were taken as independent fitting parameters.
For all films, $2\Delta(0)/k_B T_c$ is found in the range of 3.6 - 4.2, as was obtained earlier in continuous NbN films \cite {Mondol2011, Kamlapure2010,Weitzel2023}. The effect of the nanopores on the BKT transition is prominently seen in these films (See  Fig.\Ref {fig:bkt}(a)-(c)). With $J_s$ decreasing appreciably, the BKT transition is observed clearly by the distinct, relatively sharp jump of $J_s$ at $T_{BKT}$ for films with mild disorder ($T_c$ as high as 7.0 K for the\ {5} {nm} thick film). Moreover, the vortex fluctuation regime, signalled by the deviation from the BCS trend, becomes larger with decreasing $J_s$ i.e. for NP films, as also apparent from the increase in $(T_{c}^{BCS} - T_{BKT})/T_{c}^{BCS}$ reported in Table \Ref {tab:experimental_RG_table}. 


\par

To understand the observed distinct BKT jumps in relatively thick (compared to NbN films reported in \cite{Weitzel2023} showing sharp BKT transitions ) and mildly disordered NP films
we solved the BKT renormalization flow equations as given below \cite{Maccari2020,Kosterlitz1974,Nelson1977}
\begin{equation} \label{eq:BKT_flow_equations}
\begin{cases}
    \frac{dK}{dl} &= -\pi K^2 g^2 \\
    \frac{dg}{dl} &= \left(2 - \pi K\right)g
\end{cases}
\end{equation}
where the $K = J_s / T$ is the renormalized superfluid stiffness and $g = 2\pi e^{-\mu / T}$ is the vortex fugacity.
Here, the ansatz for the vortex-core energy $\mu(T) = \gamma J_s(T)$ is used, with $\gamma$ as a tuning parameter to match the result of the RG equations with the experimental data. 
Since the BCS fit is a good description of the low-temperature regime, in which vortices do not yet play a role in superfluid stiffness depletion, we used the BCS fits of the experimental data as initial conditions for the BKT RG flow so that $K(T, l=0) = J_{BCS}(T)/T$, with $J_{BCS}$ obtained from Eq.\eqref{BCSfit}.
The RG BKT fits of the experimental data (for the 5 nm thick film) 
are shown in Fig.\ref{fig:bkt}(a)-(c) for the continuous and NP films with pore diameters of 40 nm and 20 nm respectively. 
In Fig.\ref{fig:bkt}(a)-(c) the green curves in the plots show the $J_s$ renormalized via the RG BKT equations, that we integrated up to a scale $l \gg 1$. 
The RG green curves for $J_s$ exhibit a sharp drop at $T_{BKT}$, as expected for the BKT transition.  However, it can be seen that the experimental data shows a much smoother transition indicating a small broadening of the transition. This is more clearly seen from the zoomed plots shown in the insets of Fig.\ref{fig:bkt}(a)-(c).

The broadening of $J_s$ around $T_{BKT}$ has previously been studied and attributed to inhomogeneities of the film \cite{Benfatto2009, Mondol2011, Maccari2017}.
To capture this broadening within the RG calculation, the stochastic method presented in~\cite{Mondol2011} is adopted, where a random variation $\epsilon_i$, normally distributed with zero mean and standard deviation $\sigma$, is added to the initial condition (For details \textcolor{blue}{see supplementary information}).
There are now two tuning parameters $\gamma$ and $\sigma$, which are adjusted to match both the transition point and the broadening. 
The resulting $J_{s}^{inh}$ are shown as red curves in Fig. \ref{fig:bkt}(a)-(c). As one can see, with this procedure, the whole temperature dependence of the experimentally measured $J_s$ can be nicely reproduced. The fitting parameters are listed in Table \ref{tab:experimental_RG_table}, providing us with additional information on the effect of nanopore geometry on the BKT transition.  The strong suppression of $J_s$ and the increase in the vortex fluctuation regime, as apparent looking at $(T_{c}^{BCS}-T_{BKT})/T_c^{BCS}$, in NP films, distinctly emerges.
 At the same time, while the bare value of $\mu$ decreases in the presence of nanopores (also obtained from MC simulations, \textcolor{blue}{see Fig. S7 in the supplementary information}), $\mu / J_s$ significantly increases in presence of nanopores. In~\cite{Mondol2011}, a similar trend was observed by reducing the thickness of the film which in turn led to a significant increase in the inhomogeneities. It is worth noting that differently from~\cite{Mondol2011}, in NP films the ratio $\sigma/J_s$, parametrizing the inhomogeneity in the system, 
remains comparable to that of continuous films. At the same time, the absolute value of $\sigma$ decreases in the presence of nanopores. This indicates that the presence of the nanopores does not lead to the drawbacks typically observed by reducing the film thickness, thus allowing to observe a relatively sharp and distinct BKT transition.
\begin{table*}[t]
    \begin{tabular}{l|c|c|c|c|c|c}
         & $J_s(0)(K)$ & $(T_{c}^{BCS}-T_{BKT})/T_c^{BCS}$ & $\mu/J_s$ & $\mu(0) (K)$& $\sigma/J_s(0)$ & $\sigma (K)$ \\ \hline
        NoNPAM~ & 232.3 & 0.09 & 0.423 & 98.27 & 0.014 & 3.25\\ 
        20nm pore~ & 21.3 & 0.22  & 2.28 & 48.71 & 0.045 & 0.96\\ 
        40nm pore~ & 11.5 & 0.21  & 7.8 & 89.87 & 0.05 & 0.58 \\ \hline
    \end{tabular}
    \caption{Table of parameters extracted from the RG fit to the experimental data for the film with thickness of {5} {nm}.}
    \label{tab:experimental_RG_table}
\end{table*}



\par {\it Conclusions -}
 We studied the BKT transition in moderately disordered SC NbN films having a periodic array of nanopores. Films of different thickness between \ {15} {nm} - \ {5} {nm} were grown on NPAMs with two different pore diameters of \ {40} {nm} and \ {20} {nm}. BKT transitions were probed through both transport and penetration depth measurements.  The zero-temperature value of $J_s$ was observed to be considerably reduced in the NP films when compared to continuous films. The BKT transition became sharp and distinct for the NP films (thickness $\approx {5} {nm}$) which were moderately disordered. The role of the nanopores on the BKT transition was studied via Monte Carlo simulations on a 2D XY model with different nanopore geometries, demonstrating the reduction of the zero temperature value of $J_s$. By solving the RG BKT equations, with the BCS values for $J_s$ as initial conditions, we reproduced the whole temperature dependence of the superfluid stiffness, $J_s$, while extracting the value $\mu$, and the disorder strength, $\sigma$, in NP films. Our results show that the reduced surface area of the superconductor and the presence of the vortex pinning centres in these NP films are responsible for enhancing their 2D character increasing the parameter space to probe the BKT physics in SC films.
\par {\it Acknowledgements -}
SB acknowledges Department of Science and Technology(DST)-Nanomission (Grant No.: DST/NM/TUE/QM-8/2019(G)/3) and Department of Atomic Energy (DAE), Government of India. Gorakhnath Chaurasiya is thanked for some initial measurements on the NP films. All experimental measurements presented here along with the analysis were done by A. V. under the supervision of S. B. The films were grown by A. V. with initial help from J. J.. I.M has conceived the theoretical model and analysis of the experimental data. The MC simulations and fits to the experimental data have been performed by R. V. under the supervision of I. M. and J. L. All authors contributed to the discussions and writing of the manuscript. The project was conceptualized by S. B..

\bibliography{BKTnano}

\end{document}


\title{Supplementary Information\\BKT phase transition in nanoporous films of superconducting NbN}

\author{A. Verma}
\thanks{These authors contributed equally}
\affiliation{School of Physical Sciences, UM-DAE Centre for Excellence in Basic Sciences, University of Mumbai, Kalina Campus, Santacruz (E), Mumbai 400098, India}
\author{R. Vedin}
\thanks{These authors contributed equally}
\affiliation{Department of   Physics, The Royal Institute of Technology, Stockholm SE-10691, Sweden}
\author{J. Jesudasan}
\affiliation{Department of Condensed Matter Physics and Materials Science, Tata Institute of Fundamental Research, Dr. Homi Bhabha Road, Colaba, Mumbai 400005, India}
\author{J. Lidmar}
\email{jlidmar@kth.se}
\affiliation{Department of   Physics, The Royal Institute of Technology, Stockholm SE-10691, Sweden}
\author{I. Maccari} 
\email{imaccari@phys.ethz.ch}
\affiliation{Department of   Physics, The Royal Institute of Technology, Stockholm SE-10691, Sweden}
\affiliation{Institute for Theoretical Physics, ETH Zurich, CH-8093 Zurich, Switzerland}
\author{S. Bose}
\email{sangita@cbs.ac.in}
\affiliation{School of Physical Sciences, UM-DAE Centre for Excellence in Basic Sciences, University of Mumbai, Kalina Campus, Santacruz (E), Mumbai 400098, India}
\date{\today}

\maketitle

\section{Growth of Nanoporous (NP) films and measurement of disorder}
\par
NbN thin films of varying thickness (t) between\ {15} - \ {5} {nm} were grown by reactive DC magnetron sputtering of a Nb target in an Ar/$N_2$ gas mixture on free-standing nanoporous anodic aluminum oxide membranes (NPAM) obtained from InRedox.  For easy handling, the NPAMs was kept on a Si (100) substrate and glued at its corners with silver paste. We used NPAMs with holes ordered in a hexagonal pattern, whose nominal hole diameters is of \ {40}{ nm} and \ {20} {nm}, and pore periods of \ {100} {nm} and \ {50} {nm} respectively.
The deposition was done at $600^{0} {C}$ substrate temperature and the $Ar/N_2$ ratio was kept at 10:1. Under these conditions, a film with thickness of\ {25} {nm} showed a high superconducting (SC) transition temperature ($T_c$) of 14.0 K. It had been shown previously that for sputtered NbN thin films,  $T_c$ gives a measure of disorder~\cite{Chocka2008,Mondol2011}. Thus, from the obtained $T_c$  we assumed that the films used for the present study were moderately disordered\cite{Mondol2011}. 
\begin{figure}[h!]
    \centering
    \includegraphics[width=1.05\linewidth, scale=0.35]{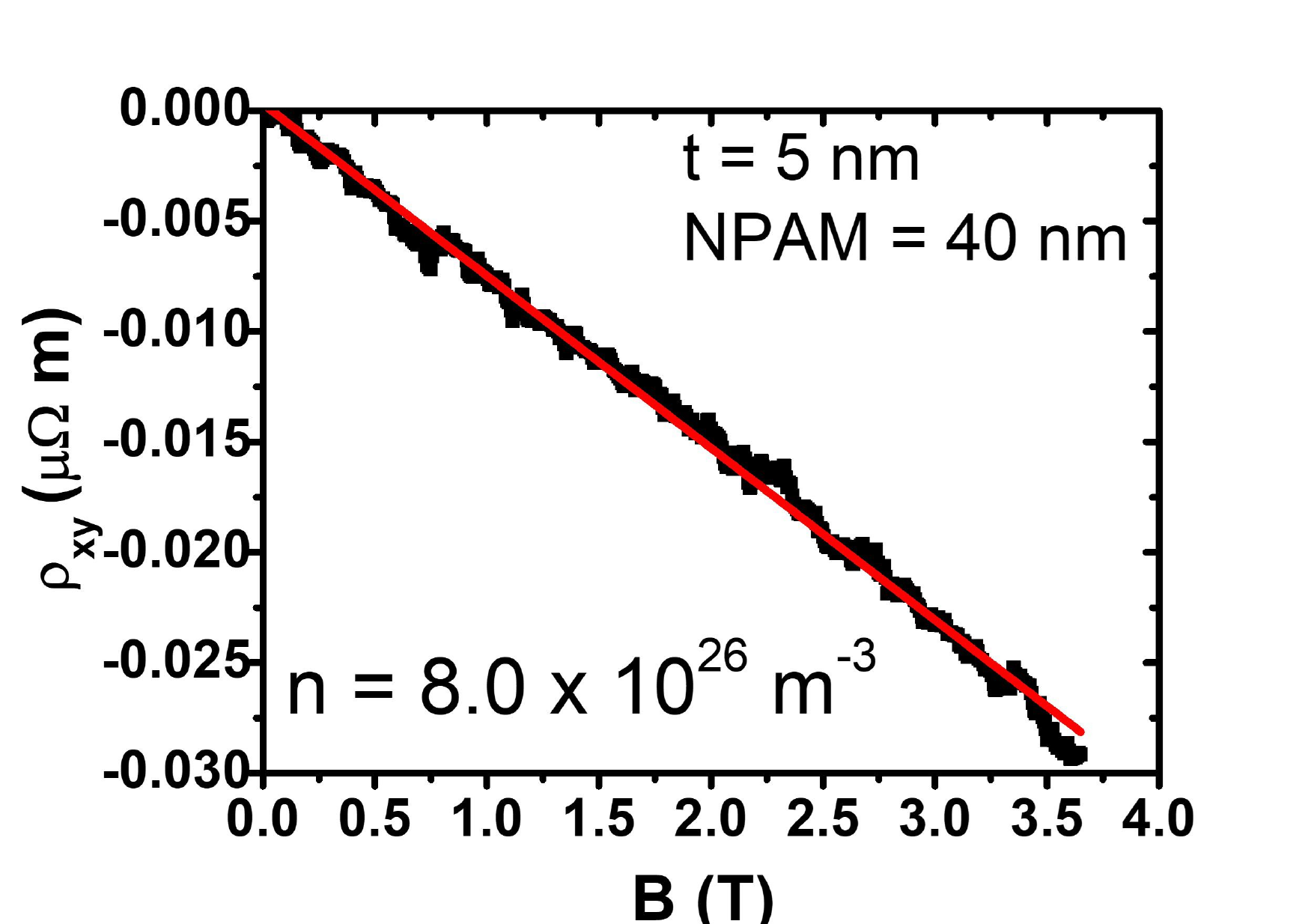}
    \caption{Hall resistivity $\rho_{xy}$ as a function of B for a NP film of thickness 5 nm measured at 20 K. The Hall coefficient $R_H$ is obtained from the slope of $\rho_{xy}$-H curve by fitting a straight line as shown. }
    \label{fig:hall}
\end{figure}
The time of deposition was varied to vary the film thickness. For the shielding response measurements, films were grown in a\ {8} {mm} circular geometry using a shadow mask during growth. For electrical transport measurements, thin strip lines,\ {0.1} {mm} wide were used in the four probe configuration. As expected, $T_c$ decreased with film thickness and did not depend substantially on the pore size. However, the $T_c$'s of the films on the NPAMs were lower when compared to continuous films. Also, the nanoporous (NP) films with circular geometry had a slightly lower $T_c$ as compared to the strip-line samples. 

While the $T_c$ of the NbN films can be taken as a measure of disorder, we carried out Hall measurements for the NP films to find the value of $k_f l$ (where, $k_f$ is the Fermi wavevector and $l$ is the mean free path) to find the level of disorder. Fig. S\Ref {fig:hall} shows the plot of the Hall resistivity as a function magnetic field, B for a NP film of thickness 5 nm grown on the NPAM with a pore diameter of 40 nm. From the Hall coefficient along with the normal state resistivity value, $k_f l$ was estimated using the Drude's model which was $\approx {4 - 5 }$ indicating that the studied films were mildly disordered.

\section{BKT transition from transport measurements for NP films}
\par
\begin{figure}[t!]
    \centering
    \includegraphics[width=1\linewidth, scale=0.45]{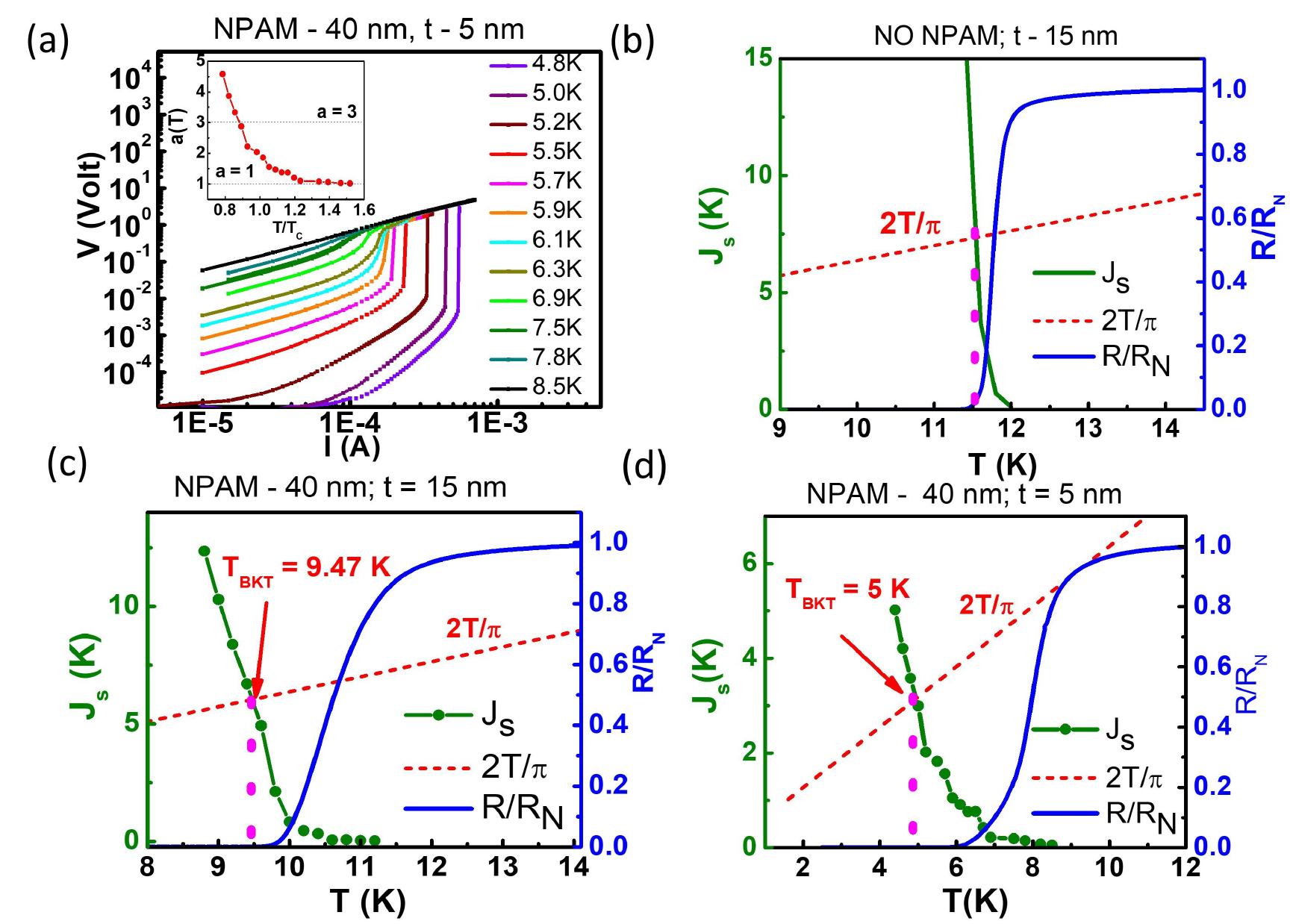}
    \caption{(a) I-V measurements on a 5 nm thick NP film grown on the NPAM with pores size of 40 nm. I and V are plotted in log scale. The inset shows the temperature variation of the extracted value of a(T) for the same film. (b) - (d) Temperature dependence of $J_s$, obtained from Eq. \eqref{eq:IV_exponent_a} along with the normalized resistance (scale on right axis) for a continuous 15 nm thick film and two NP films with pore diameter of \ {40} {nm} and thickness of \ {15} and \ {5} {nm} respectively. The universal $\frac {2T} {\pi}$ BKT line is also shown in red. }
    \label{fig:jstrans}
\end{figure}

To check the BKT transition of the NP films, current-voltage (I-V) measurements were carried out at temperatures below their $T_c$. A representative plot of the I-V measurement carried out on a 5 nm thick NP film on a NPAM of pore size of 40 nm is shown in Fig. S\ref{fig:jstrans}(a), both I and V are shown in log scale. As is known, according with the BKT transition, from the Nelson-Halperin scaling theory~\cite{Nelson1977, Venditti2019}, the I-V characteristics acquire a nonlinear dependence near $T_c$, where, $V \propto I ^ {a(T)}$. Here the exponent is related to $J_s$ as, 
\begin{equation}\label{eq:IV_exponent_a}
a(T) = 1 + \frac {\pi J_s (T)} {T} .
\end{equation}
A representative plot of the temperature variation of $a(T)$ extracted from the I-V measurements is shown as an inset of Fig.~S\Ref {fig:jstrans}(a). At the critical temperature, $T_{BKT}$, ideally, $a(T)$ should jump from 3 to 1. That is a consequence of the discontinuous jump of the superfluid stiffness, $J_s(T )$, from $J_s(T^-_{BKT}) = \frac{2T^-_{BKT}}{\pi}$ to $J_s(T^+_{BKT})=0$.
However, in real 2D superconductors, $J_s$ is seen to go continuously to zero. In Fig.~S\Ref {fig:jstrans}(b)-(d), the variation of $J_s$ with temperature is shown for a 15 nm thick continuous film along with two of the NP films with pore diameter of\ {40} {nm} and thicknesses of\ {15} {nm} and \ {5} {nm} respectively. In the same plot, the universal $T_{BKT}$ line is shown in red and its intersection with $J_s$ is taken as $T_{BKT}$. As discussed previously ~\cite{Benfatto2009,Venditti2019}, the low vortex-core energy and a moderate inhomogeneity of the film account for the smearing of the BKT jump. However, what seems surprising is that for films with thickness $t$ larger than the coherence length $\xi$, small yet visible signatures of BKT transition can be observed ($T_{BKT}$ well seperated from $T_c$)(Note: From Fig. S\ref{fig:jstrans}(b), for a 15 nm thick continuous film, $T_{BKT}$ is same as $T_c$ ).  This can be attributed to the geometry of the films where the pore diameter is still larger than both $t$ and $\xi$ making them behave as quasi-2D in this regime. 
\section{BKT transition from penetration depth measurements for NP films on NPAM}

\begin{figure}[t!]
     \includegraphics[width=1.2\linewidth]{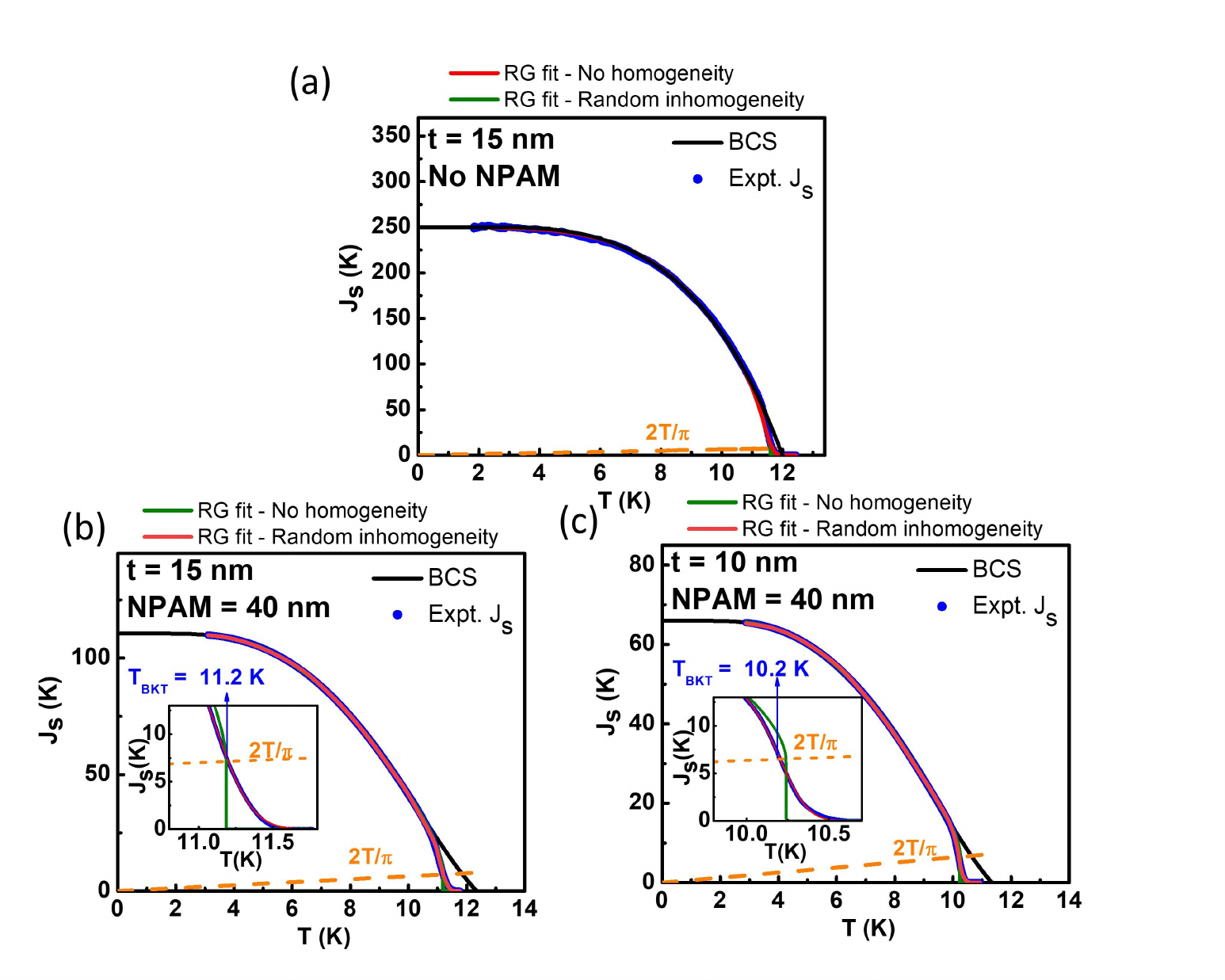}
    \caption{(a) - (c) Temperature variation of $J_s$ extracted from penetration depth measurements for continuous 15 nm thick film and NP films grown on NPAM with pore diameter of \ {40} {nm} with thickness of \ {15} {nm} and \ {10} {nm} respectively. }
    \label{fig:40nm}
\end{figure}

\begin{figure}[h!]
 \centering
    \includegraphics[width = 1.1\linewidth]{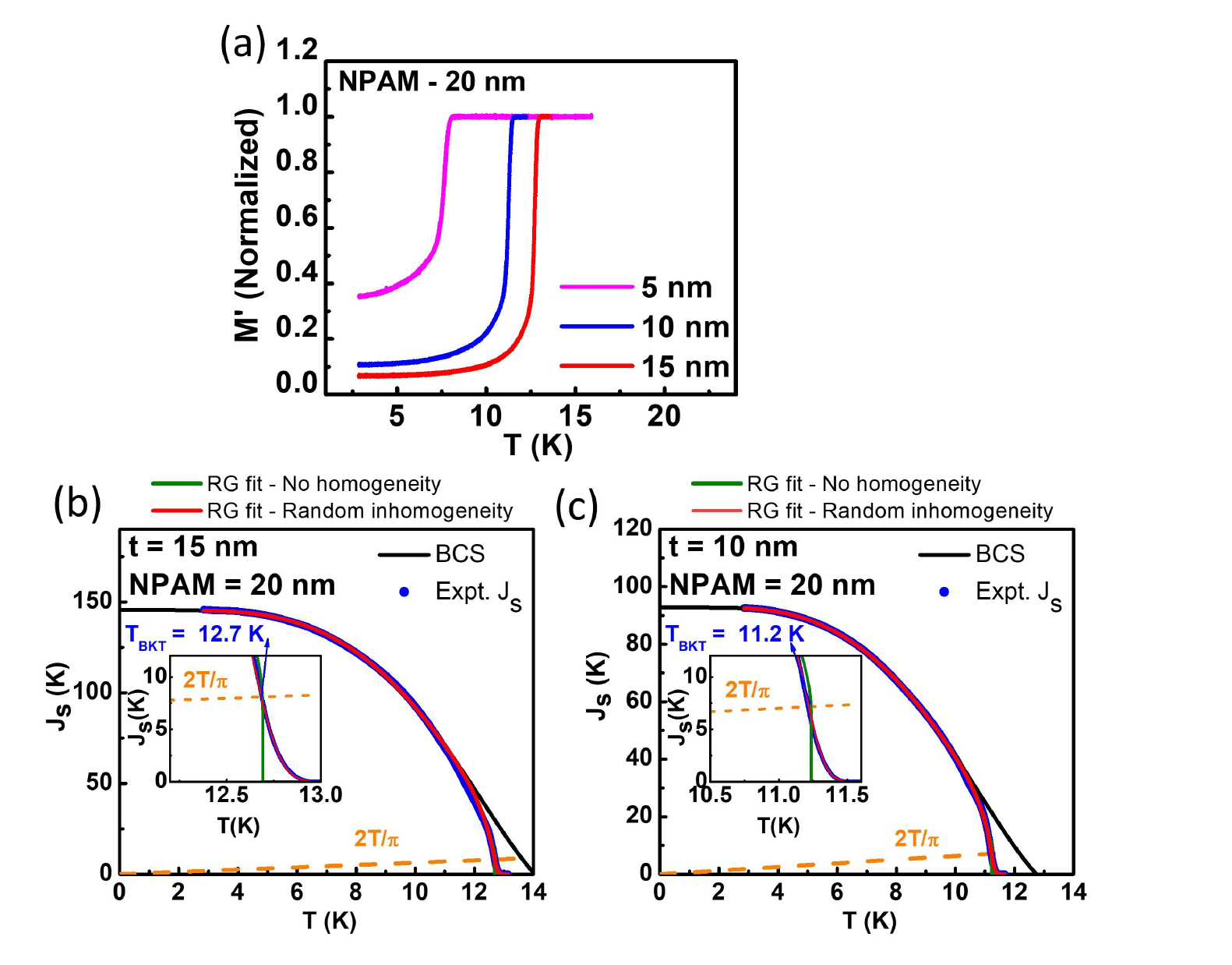}
    \caption{ (a) Real part of mutual inductance ($M{'}$) $\it {vs}$ T for films with thickness varying between\ {15} and \ {5} {nm} on the NPAM with pore diameter of\ {20} {nm.} (b) - (c) Temperature variation of  $J_s$ extracted from penetration depth measurements for NP films grown on the NPAM with pore diameter of \ {20} {nm} and thickness of 15 nm and 10 nm respectively. }
    \label{fig:20nm}
\end{figure}
The temperature variation of $J_s$ was obtained from penetration depth measurements using the analysis described in Ref. \cite{Gupta2020}. Fig. S\Ref {fig:40nm}(a) shows the variation for a 15 nm continuous film. As expected no BKT transition is observed. In Figs. S\Ref {fig:40nm}(b)-(c), the variation of $J_s$ is shown for the NP film grown on the NPAM with a pore diameter of 40 nm with a thickness of 15 nm and 10 nm respectively. Interestingly, for both films, a small signature of the BKT transition can be seen which gets slightly more pronounced for the 10 nm thick film. This is similar to that obtained from transport measurements confirming that NP films enhance the 2D character of the films. Fig. S\Ref {fig:20nm}(b)-(c) shows the same variation for the NP films grown on the NPAM with pore diameter of 20 nm for film thickness of 15 nm and 10 nm respectively. These have been obtained from analyzing the penetration depth data measured using the two-coil mutual inductance technique (M' vs T plots for the three films is shown in Fig. S\Ref {fig:20nm}(a)). Here, too, we observe the same characteristic feature of the BKT transition becoming more pronounced with decreasing thickness or decreasing $J_s$. For the lowest film thickness studied (5 nm), a sharp, distinct jump at $T_{BKT}$ is seen, very similar to the 5 nm film grown on the 40 nm diameter NPAM (See main text).

\begin{figure}[b]
 \centering
    \includegraphics[width =0.8 \linewidth]{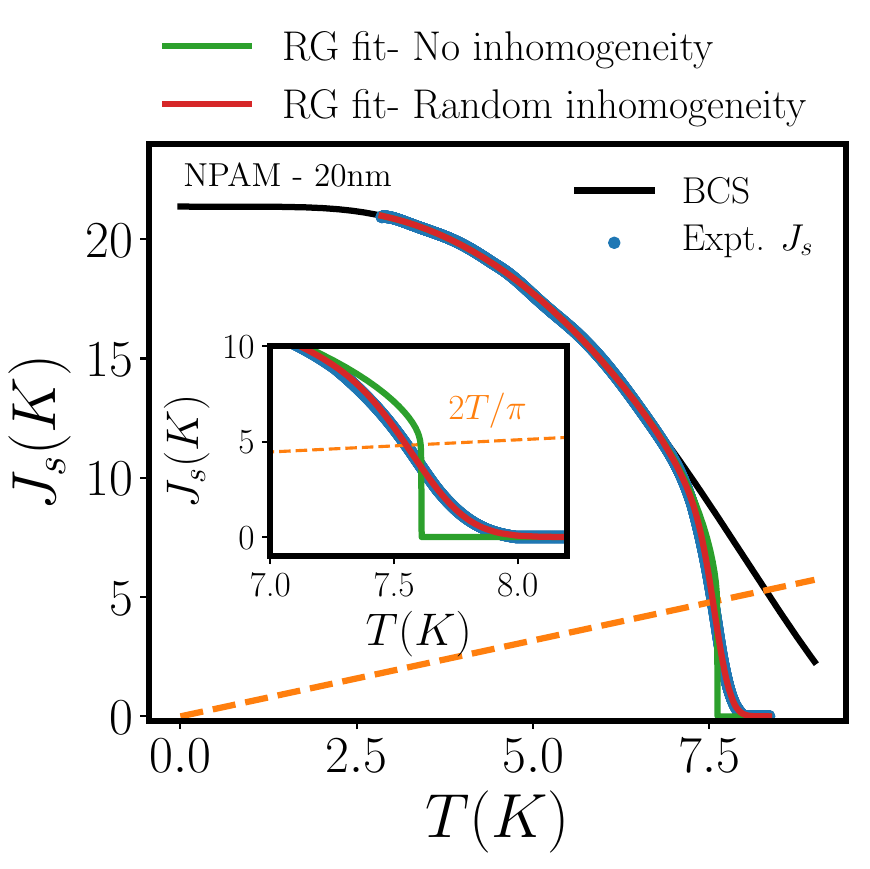}
    \caption{ Temperature dependence of (blue) the measured stiffness for the {5} {nm} thick film with {20} {nm} diameter pores. The fitted stiffness calculated using the BKT RG equations (green) without random inhomogeneity, and (red) with random inhomogeneity. The inset shows a zoomed-in view close to $T_{BKT}$.}
    \label{fig:inhom_RG_fit_20nm_pores}
\end{figure}

\section{Fitting experimental data with BKT renormalization group equations}
To describe the full temperature dependence of the experimental measurements, we analyzed the superfluid stiffness using a two-step renormalization group (RG) approach~\cite{Maccari2020}.
This approach consists of solving the BKT flow equations~\cite{Kosterlitz1974,Nelson1977} for the renormalized stiffness $K = J_s/T$ and the vortex fugacity $g = 2\pi e^{-\mu/T}$
\begin{equation}
    \begin{cases}
        \frac{d K}{d \ell} &= -\pi K^2 g^2 \\
        \frac{d g}{d \ell} &= (2 - \pi K)g
    \end{cases}
        \label{eq:BKT_flow_equations}
\end{equation}
where the dimensionless length $\ell=\log(L/L_0)$, $L$ the linear size of the film, and $L_0$ a characteristic length scale. We assume $L_0 \ll L$ and the equations are integrated up to a $\ell_{max} \gg 1$.
We take as initial condition for $K$ a BCS fit to the experimental stiffness for lower temperatures
\begin{equation}
\begin{cases}
    K(\ell=0) &= \frac{J_{BCS}}{T} \\
    g(\ell=0) &= 2\pi e^{-\mu / T}
\end{cases}
\end{equation}
and for the vortex-core energy we make the ansatz $\mu = \gamma J_s$.

The measured stiffness for the NP film grown on NPAM with pores of diameter {20} {nm} is shown in Fig. S\ref{fig:inhom_RG_fit_20nm_pores}, with the BCS fit shown as black line.
The green curve in the plot shows instead the superfluid stiffness renormalized via the RG flow Eq.~\eqref{eq:BKT_flow_equations}.
The RG green curve for $J_s$ exhibits a sharp drop at $T_{BKT}$, as expected for the BKT transition, yet the experimental data show a much smoother transition.

The broadening of $J_s$ around $T_{BKT}$ has previously been studied and attributed to inhomogeneities of the film \cite{Benfatto2009, Mondol2011,Maccari2017}.
To capture this broadening within the RG calculation, we follow a stochastic method presented in~\cite{Mondol2011}, where a random variation $\epsilon_i$ is added to the initial condition
\begin{equation}
\begin{cases}
    K_i(\ell=0) &= \frac{J_{BCS} + \epsilon_i}{T} \\
    g_i(\ell=0) &= 2\pi e^{-\mu_i / T}
\end{cases}
\end{equation}
with $\mu_i = \gamma T K_i$. This random variation $\epsilon_i$ is taken to be normally distributed with zero mean and standard deviation $\sigma$.

This gives us two tuning parameters $\gamma$ and $\sigma$, which are adjusted to match both the $T_{BKT}$ and the broadening of the transition.
With the renormalized stiffness integrated up to $l_{max} \gg 1$ the result is obtained as the mean value
\begin{equation}
    J_{s}^{inh} = \sum\limits_{i=1}^{N} \Delta \epsilon_i P(\epsilon_i) T K_i,
\end{equation}
where the probability of the variation $P(\epsilon_i) = \frac{1}{\sqrt{2 \pi \sigma^2}}e^{-\epsilon_i^2 / 2 \sigma^2}$.
The resulting $J_{s}^{inh}$ is shown as the red curve in Fig. S\ref{fig:inhom_RG_fit_20nm_pores}. As one can see, with this procedure, we can nicely reproduce the whole temperature dependence of the superfluid stiffness experimentally measured. The fitting parameters are collected in Table 1 of the main text, providing us with additional information on the effect of different nanopore geometries on the BKT transition.

\section{Fitting Monte Carlo simulation data With BKT renormalization group equations}
\par

\begin{figure*}[t!]
 \centering
    \includegraphics[width = \textwidth]{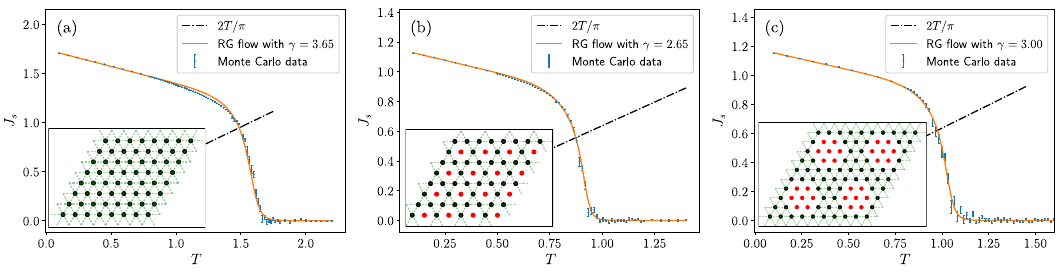}
    \caption{Simulated stiffness fitted with the BKT RG equations for (a) a lattice without pores with size $L=108$, (b) smaller pores close together S=1, W=1 with size $L=108$, and (c) larger pores farther apart S=7, W=2 with size $L=105$. The insets shows a small section of the lattice geometry.}
    \label{fig:mc_RG_fits}
\end{figure*}

The superfluid stiffness $J_s$ signals the onset of a SC diamagnetic response in the system, as such it is defined as the system's response to a phase twist $\Theta$.  For a given direction, say along the x-axis, it is calculated as
\begin{equation}
    J_s^x = \frac{1}{\Omega} \cdot \frac{\partial^2F}{\partial \Theta_x^2}\Big|_{\Theta_x = 0} = J_d^x - J_p^x,
\end{equation}
where $\Omega= \frac{\sqrt{3}}{2} N_\mathrm{SC}$ is the area of the SC region, with $N_\mathrm{SC}$ the number of SC sites, and $F$ is the free energy of the system, $F = -k_B T \ln{Z}$. The $J_d^x$ and  $J_p^x$ are respectively the diamagnetic and paramagnetic contributions to $J_s^x$, whose explicit equation is
\begin{equation}
    J_d^x = \frac{1}{\Omega}\left\langle \sum\limits_{<ij>} J_{i,j} (\vec{x}\cdot \vec{e}_{ij})^2 \cos(\theta_j - \theta_i) \right\rangle,
\end{equation}
and 
\begin{equation}
    J_p^x = \frac{1}{\Omega k_B T} \left\langle \left[ \sum\limits_{<ij>} J_{i,j} \left( \vec{x}\cdot \vec{e}_{ij} \right) \sin(\theta_j - \theta_i) \right]^2 \right\rangle, 
\end{equation}
where $\vec{e}_{ij}$ is the lattice vector connecting two nearest-neighbours sites, $i$ and $j$, and $\braket{\dots}$ indicates the thermal average. That is obtained averaging over 50000 MC steps, once having discarded the first 5000 needed for the system to thermalize. The error bars are computed with a Bootstrap resampling method, where the data is first binned into blocks of size 500 MC steps in order to account for the finite correlation time in the MC data. 
\par
As with the experimental data the vortex-core energy $\mu$ is anlyzed for the simulated nanopore lattice geometries using the RG flow Eq. \eqref{eq:BKT_flow_equations}.
Since the RG flow only accounts for topological excitations of the phase, we take as initial conditions the low-temperature dependence of the simulated $J_s(T)$, which are due to the non-topological spin-wave excitations. These Gaussian fluctuations renormalize the stiffness linearly, such that $J_s^{SW}(T) = J_s(0) - \alpha_{SW}T$. The $J_s(0)$ and $\alpha_{SW}$ are obtained by fitting the simulated $J_s$ in the low-temperature regime, and for the vortex-core energy the ansatz $\mu = \gamma J_s$.
\begin{equation}
    \begin{cases}
        K(\ell=0) &= \frac{1}{T}[J_s(0) - \alpha_{SW} T] \\
        g(\ell=0) &= 2\pi e^{-\gamma K(l=0)}
    \end{cases}
    \label{eq:initial_cond}
\end{equation}
Because of the finite lattice sizes $L$ we integrate the RG equations up to a $\ell_{\max} = \log(L/L_0)$. In this way, there are 2 parameters, $\gamma$ and $L_0$, which are tuned to fit the result of the RG calculation to the simulated stiffness.
The RG fits for 3 different configurations, the clean lattice without pores, smaller pores close together (S=1, W=1), and larger pores farther apart (S=7, W=2) are shown in Fig. S\ref{fig:mc_RG_fits}(a)-(c), while the values of $J_s(0)$, $T_{BKT}$, and $\gamma$ are collected in Table S\ref{tab:mc_RG_table}. 

\begin{figure}[h!]
 \centering
    \includegraphics[width = 0.8\linewidth]{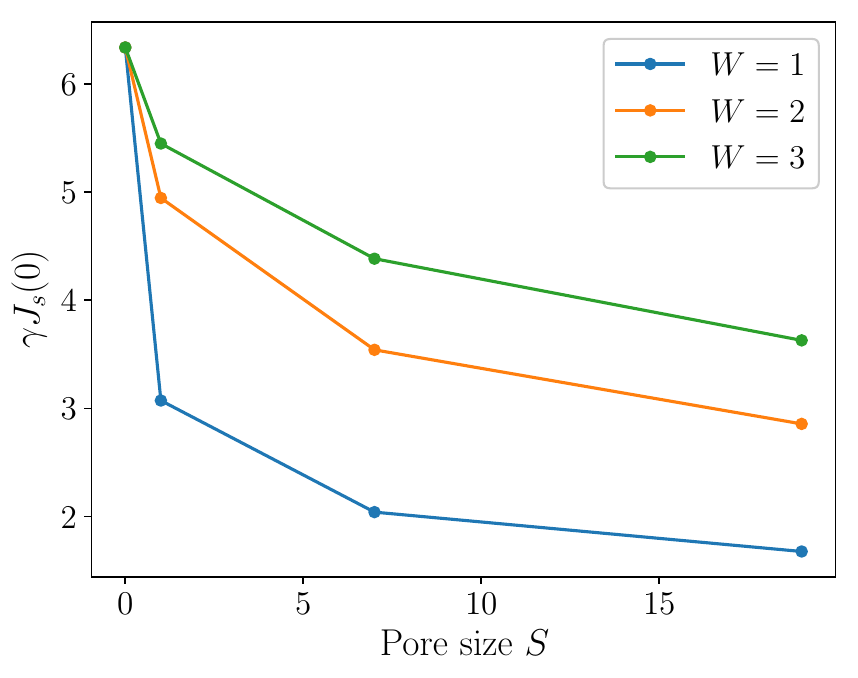}
    \caption{Vortex-core energies $\mu = \gamma J_s$ as calculated by fitting the MC data with the RG flow equations for each pore size $S=1,7,19$ and pore separation $W=1,2,3$.}
    \label{fig:mc_vortex_core_energies}
\end{figure}

As shown in Fig. S\ref{fig:mc_vortex_core_energies} the vortex-core energy $\mu=\gamma J_s(0)$ decreases with increasing pore size. At the same time, similarly to what has been observed experimentally, $\gamma$ increases when increasing both the pore size and the pore separation.
The two main discrepancies between the MC numerical results and the experimental observations are the following. On the one hand, the absence of a comparable depletion of $J_s(0)$ when nanopores are added to the system. 
On the other, unlike the experimental data, the clean lattice without pores is found to have the largest value of $\gamma$. 
Note also that, in our MC simulations, the increase of $\gamma$ is not accompanied by a corresponding decrease of $J_s(0)$ and $T_{BKT}$, see Table~S\ref{tab:mc_RG_table}.

\begin{table}[h]
    \centering
    \begin{tabular}{l|c|c|c|c|c}
          & $~J_s(0)~$ & $~T_{BKT}~$ & $~\mu/J_s~$ & $~\mu(0)~$ & $~L_0~$ \\ \hline
         Clean~ & 1.74 & 1.50 & 3.65 & 6.33 & 7.5\\
         S=1, W=1~ & 1.16 & 0.88 & 2.65 & 3.07 & 8.5 \\
         S=7, W=2~ & 1.19 & 0.98 & 3.0 & 3.54 & 7.5 \\ \hline
    \end{tabular}
    \caption{Table of parameters extracted from the RG fit to the Monte Carlo data.}
    \label{tab:mc_RG_table}
\end{table}

\bibliography{BKTnanosupple}